# Estereoscópio com tela holográfica para ver tomografias

**José J. Lunazzi*, Rolando L. Serra** and Daniel S. F. Magalhães***

*Laboratório de Óptica, Instituto de Física Gleb Wataghin, P.O.Box 6165, University of Campinas - UNICAMP, 13083-970 Campinas, SP, Brazil

**Departamento de Física, Instituto Superior Politécnico "José Antonio Echeverría" (Cujae), Ave. 114, 11901, Marianao, Ciudad de La Habana, CP 19390, Cuba, Brazil

**RESUMO**

*A estereoscopia é uma técnica que permite a observação de imagens tridimensionais, mas está sempre associada com o uso de algum equipamento especial para visualização, como óculos bicolores ou polarizados. Para o uso em aplicações médicas o emprego de tais equipamentos pode inviabilizar sua utilização durante procedimentos cirúrgicos, por exemplo.*

*Neste trabalho apresentamos um novo tipo de estereoscópio que utiliza uma tela holográfica para geração de imagens tridimensionais sem o uso de qualquer equipamento adicional. Apresentamos a descrição do equipamento utilizado e resultados das imagens visualizadas.*

**INTRODUÇÃO**

A estereoscopia permite ver imagens em terceira dimensão completa, é uma técnica tão antiga quase quanto à fotografia e tem aplicações na fotografia, cinema [1], televisão [2], e em técnicas de levantamento geodésico, por exemplo. Necessita sempre de óculos ou visores especiais que limitam aos movimentos da pessoa que deseja alternar seu uso com o de telas convencionais, como pode ser o caso de um cirurgião em sala de cirurgia ou um operador de tráfego aéreo. Eliminar a necessidade de usar os óculos torna, por sua vez, a visualização para um número grande de pessoas mais higiênica porque dispensa o uso do mesmo óculos para diferentes pessoas, ou barata, porque evita ter de dar um óculos a cada observador.

Temos conseguido isto por meio da tela holográfica [3], elemento que vem sendo desenvolvida desde o ano de 1987 [4-6]. Depois que a tela holográfica foi patenteada e exibida em vários congressos internacionais em diversos paises, outros projetos de originalidade duvidosa descrevem o elemento e aplicações da mesma [7,8]. Na primeira versão que desenvolvemos não realizávamos a combinação de comprimentos de onda de maneira a preservar a cor, e agora o fazemos com bastante sucesso, e se algum defeito de cor aparece em nosso protótipo atual deve-se a que ainda não o fizemos com otimização de parâmetros. Mas isto não é importante no caso de tomografias, onde a cor nunca é original e somente usa-se pseudo-coloração para destacar regiões ou tonalidades.

**DESCRIÇÃO**

Uma tela holográfica projeta na posição do observador uma distribuição de luz que é a imagem holográfica de um difusor colocado na hora da tomada da tela. Na figura 1 mostramos como a tela holográfica coloca essa imagem deslocada em função do valor do comprimento de onda. Escolhendo do espectro contínuo somente três cores para exemplificar, deixamos a idéia da presença das cores intermediárias para o leitor.

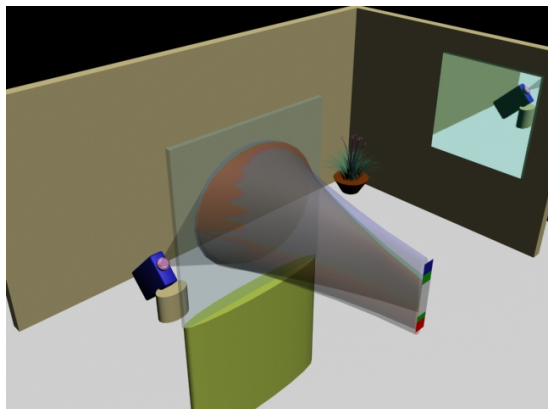

*Figura 1: Posição das imagens do difusor em função do comprimento de onda. A iluminação vem por baixo.*

Na figura as imagens se sobrepõem somente de maneira parcial para melhor ilustrar o caso, e permite entender que, dando ao difusor maior extensão vertical, haverá uma região onde todos os comprimentos de onda vão chegar sobrepostos e nela teremos a preservação da cor original. Quando esta tela é realizada com um difusor amplo tanto na horizontal como na vertical temos uma extensa região para observar a figura, mas cada olho vê sempre a mesma e é o caso da tela holográfica 2D, que hoje chama-se simplesmente de "tela holográfica para projeção" porque a tela holográfica 3D é invenção nossa desconhecida do público ainda. Esse tipo de tela começa a ser usado no Brasil [9]. Para uso em estereoscopia deve-se limitar o campo horizontal de maneira a não ser muito maior que a distância interpupilar de uma pessoa, porque a região de aproveitamento da luz não vai ser maior do que isso.

Vemos na figura 2 como devem ser colocados os dois projetores em incidência vertical, ou seja, por baixo ou por cima da tela, para que cada olho do observador receba a imagem vinda de cada projetor isoladamente.

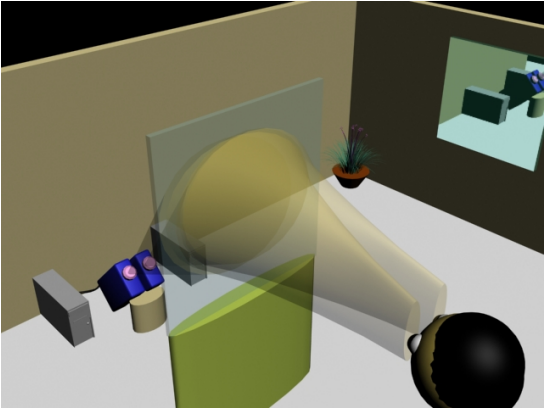

*Figura 2: Projeção sobreposta na tela de dois projetores, cada um gerando um campo de visão diferente na posição do observador.*

É a diferença de posição lateral dos projetores que gera diferentes posições laterais para o campo do observador. O fato de a incidência ser obliqua gera o efeito trapézio, hoje corrigível facilmente pelos projetores, mas a obliqüidade lateral gera um efeito de distorção diferenciado que é pequeno e temos desprezado em um primeiro momento.

**Alguns parâmetros fundamentais do processo**

Porque o tamanho da mesa holográfica de que dispomos é limitado não se consegue aproximar mais a posição do observador respeito da tela. Se o filme holográfico fosse de até 1.500 linhas /mm de resolução, seria possível usar uma geometria mais favorável, mas o único filme holográfico que é fabricado hoje resolve mais de 3.000 linhas por milímetro, e obriga a gravar a tela iluminando pelo mesmo lado. São características holográficas que limitam a proximidade a que se pode colocar o difusor sem gerar sombra no filme holográfico. A projeção acontece também em direção oposta à do feixe de referencia no registro e tem, portanto, curvatura diferente à ideal, que seria a mesma do feixe de referência. Isto também age negativamente afastando a imagem do difusor da tela, obrigando a colocar ao observador mais distante. Os projetores não podem então estar muito perto da tela, e como não contam com lentes de distância focal suficiente, boa parte da imagem fica sobrando por fora da tela, perdendo em brilho e resolução. Para estas limitações existe procedimento de construção da tela que as eliminaria, e projetores preparados para projetar a maiores distâncias, o que esperamos poder implementar no futuro. O resultado que temos de momento, com o observador a 2 m da tela, é já um resultado com aplicações interessantes.

**RESULTADOS EXPERIMENTAIS**

Temos construído uma tela holográfica de 40 cm x 40 cm com eficiência de difração no vermelho de 3,5 % e largura de área para visão do observador de 2 cm. A altura foi de 1,2m cm e projetamos por meio de dois projetores marca Viewsonic modelo PJ503D, cada um com um componente de par estéreo tomográfico.

Ilustramos na figura 3 o par estéreo de uma tomografia de crânio, como é visto na tela por cada olho.

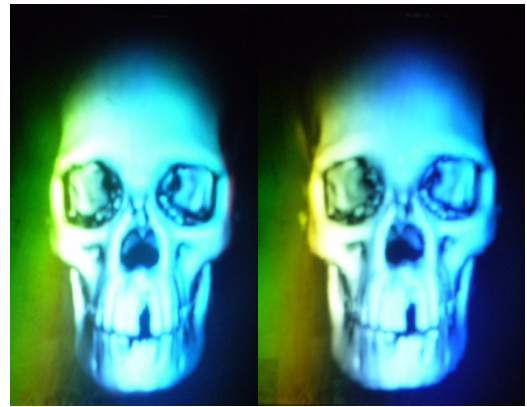

*Figura 3: esq., vista da tela pelo olho esquerdo. Dir., vista da tela pelo olho direito.*

Nossa figura é monocromática, a possibilidade de cor nela vem do fato dela estar perto da cor branca original. A visão resultante é equivalente à de um bom estereoscópio, considerando a resolução limitada de que dispomos no momento.

**CONCLUSÃO**

Com uma tela holográfica especialmente desenvolvida visando aplicações estereoscópicas foi possível obter um sistema gerador de imagens tridimensionais. Esse sistema pode ser empregado para visualização tridimensional de tomografias sem o uso de equipamentos especiais. A limitação do sistema é o pequeno no campo horizontal de visualização do observador.